\documentclass[a4paper,UKenglish,cleveref, autoref, thm-restate]{oasics-v2021}

\usepackage[table,xcdraw]{xcolor}
\usepackage{booktabs}
\usepackage{array}
\usepackage{tabularx}

\usepackage[nolist]{acronym}
\usepackage{amsmath}
\usepackage{amsfonts}
\usepackage{algorithm}
\usepackage{algorithmic}
\usepackage{wasysym} 
\usepackage{adjustbox} 

\title{H-MBR: Hypervisor-level Memory Bandwidth Reservation for Mixed Criticality Systems} 
\newcommand{\priority}[2]{%
    \ifnum#1=100\cellcolor{#2}\checkmark\else\fi%
}
\titlerunning{H-MBR: Hypervisor-level Memory Bandwidth Reservation for MCS}

\author{Afonso Oliveira}{Centro ALGORITMI / LASI, Universidade do Minho, Portugal}{pg53599@alunos.uminho.pt}{https://orcid.org/0009-0007-9996-940X}{}

\author{Diogo {Costa}}{Centro ALGORITMI / LASI, Universidade do Minho, Portugal}{diogocostaes21@gmail.com}{https://orcid.org/0000-0002-0725-5861}{Supported by FCT grant 2022.13378.BD}

\author{Gonçalo Moreira}{Centro ALGORITMI / LASI, Universidade do Minho, Portugal}{pg53841@alunos.uminho.pt}{https://orcid.org/0009-0009-9883-1403}{}

\author{José Martins}{Centro ALGORITMI / LASI, Universidade do Minho, Portugal}{jose.martins@dei.uminho.pt}{https://orcid.org/0000-0001-9380-7150}{Supported by FCT grant SFRH/BD/138660/2018}

\author{Sandro Pinto}{Centro ALGORITMI / LASI, Universidade do Minho, Portugal}{sandro.pinto@dei.uminho.pt}{https://orcid.org/0000-0003-4580-7484}{}

\authorrunning{A. Oliveira et al.} 

\Copyright{Afonso Oliveira, Diogo Costa, Gonçalo Moreira, Jose Martins, Sandro Pinto} 



\category{} 

\relatedversion{} 

\EventEditors{Patrick Meumeu Yomsi and Stefan Wildermann}
\EventNoEds{2}
\EventLongTitle{Sixth Workshop on Next Generation Real-Time Embedded Systems (NG-RES 2025)}
\EventShortTitle{NG-RES 2025}
\EventAcronym{NG-RES}
\EventYear{2025}
\EventDate{January 20, 2025}
\EventLocation{Barcelona, Spain}
\EventLogo{}
\SeriesVolume{128}
\ArticleNo{5}

\begin{document}

\maketitle


\begin{abstract} 
Recent advancements in fields such as automotive and aerospace have driven a growing demand for robust computational resources. Applications that were once designed for basic Microcontroller Units (MCUs) are now deployed on highly heterogeneous System-on-Chip (SoC) platforms. While these platforms deliver the necessary computational performance, they also present challenges related to resource sharing and predictability. These challenges are particularly pronounced when consolidating safety-critical and non-safety-critical systems, the so-called Mixed-Criticality Systems (MCS) to adhere to strict Size, Weight, Power, and Cost (SWaP-C) requirements.
 MCS consolidation on shared platforms requires stringent spatial and temporal isolation to comply with functional safety standards (e.g., ISO 26262). Virtualization, mainly leveraged by hypervisors, is a key technology that ensures spatial isolation across multiple OSes and applications; however ensuring temporal isolation remains challenging due to contention on shared resources, such as main memory, caches, and system buses, which impacts real-time performance and predictability. To mitigate this problem, several strategies (e.g., cache coloring and memory bandwidth reservation) have been proposed. 
Although cache coloring is typically implemented on state-of-the-art hypervisors, memory bandwidth reservation approaches are commonly implemented at the Linux kernel level or rely on dedicated hardware and typically do not consider the concept of Virtual Machines that can run different OSes. To fill the gap between current memory bandwidth reservation solutions and the deployment of MCSs that operate on a hypervisor, this work introduces \textit{H-MBR}, an open-source VM-centric memory bandwidth reservation mechanism. \textit{H-MBR} features (i) VM-centric bandwidth reservation, (ii) OS and platform agnosticism, and (iii) reduced overhead. Empirical results evidenced no overhead on non-regulated workloads, and negligible overhead (<1\%) for regulated workloads for regulation periods of 2 µs or higher.
\end{abstract}

\section{Introduction}
\label{sec:introduction}

Rapid advancements in industries such as automotive, aerospace, and industrial automation have led to increasingly demanding applications, driving a significant need for higher computational power \cite{cerrolaza2020multi, burgio2017automotive}. These applications now require systems capable of handling a diverse range of computationally intensive tasks, such as autonomous driving, flight control, and complex industrial automation workflows \cite{gracioli2019}. To address this computational needs, the once simple microcontroller units (MCUs) evolved to sophisticated heterogeneous architectures. Modern designs integrate multiple Central Processing Units (CPUs) alongside specialized accelerators, such as Graphics Processing Units (GPUs)\cite{burgio2017automotive}, Field-Programmable Gate Arrays (FPGAs)\cite{gracioli2019}, and novel AI accelerators like Tensor Processing Units (TPUs) and Neural Processing Units (NPUs)\cite{lai2018cmsis,PULP-NN}. This shift enables efficient processing of complex workloads but simultaneously introduces challenges related to resource sharing and system predictability, especially as systems grow in complexity and criticality\cite{reder2020interference}.
In addition to computational demands, embedded systems increasingly face Size, Weight, Power, and Cost (SWAP-C) constraints, which are driving a trend towards system consolidation\cite{cerrolaza2020multi,garcia2014challenges}, giving rise to the so called Mixed-Criticality Systems (MCS). However, integrating MCS on shared platforms faces a significant challenge: stringent requirements for both spatial and temporal isolation \cite{yun2014palloc, gracioli2019, abella2015wcet, mancuso2013cache, farshchi2020bru}. These systems must adhere to certification standards (e.g., ISO 26262 in automotive, CENELEC for railway systems, and ECSS for space applications) which place strict demands on system safety, reliability, and predictability \cite{modica2018hypervisor, casini2020holistic, crespo2018feedback, serrano2021qos, jailhouse}.

Virtualization has become essential in the consolidation of MCS, with hypervisors playing a pivotal role. Hypervisors, particularly static partitioning hypervisors, must be minimal yet sufficiently robust to ensure reliable isolation while meeting the real-time (RT) performance requirements of modern applications \cite{martins2020bao, crespo2018feedback, pinto2019trustzone, oliveira2024ia}. Despite advancements in spatial isolation through various partiotioning techniques, temporal isolation remains an active area of research due to contention on shared resources \cite{ye2014coloris, yun2014palloc, modica2018hypervisor}. These shared resources, including (i) main memory, (ii) caches, and (iii) the system bus, can introduce substantial unpredictability in response times if left unregulated \cite{casini2020holistic, reder2020interference, farshchi2020bru}. Such variability in access latency poses significant risks to RT, safety-critical applications, where predictable performance is essential to meet strict timing constraints \cite{guthaus2001mibench, pellizzoni2013predictable, jailhouse}. To address shared resource contention, considerable efforts have been made in both academia and industry. Techniques such as (i) cache coloring and (ii) MBR have emerged as widely recognized approaches to regulate and reduce contention induced by memory access \cite{yun2014palloc, ye2014coloris, mancuso2013cache}. However, unlike cache coloring, most MBR approaches are implemented at the OS-level, typically on the Linux kernel \cite{farshchi2020bru, modica2018hypervisor}, narrowing their flexibility and applicability, underscoring the need for a more versatile solution, one that can provide robust temporal isolation across diverse applications and platforms without sacrificing configurability or performance \cite{cazorla2019probabilistic, kim2017cache, casini2020holistic}.

In this paper we present H-MBR, an open-source VM-centric MBR mechanism implemented in Bao hypervisor. Its primary contributions over traditional MBR methods stem from an innovative design: (i) VM-Centric bandwidth allocation, (ii) OS and platform agnosticism, and (iii) reduced overhead. Furthermore, H-MBR supports unbalanced distribution of bandwidth across a VM's vCPUs. As H-MBR's is implemented at the hypervisor level it enables the memory bandwidth regulation of VMs with different operating systems, (e.g., Linux, FreeRTOS and Zephyr). Additionally, H-MBR is designed to be platform-agnostic, allowing it to be ported across various architectures such as ARMv8-A, ARMv8-R, and RISC-V . Finally, as Bao hypervisor stands as the hypervisor with lowest interrupt latency among the state-of-the-art static partitioning hypervisors \cite{martins2023static}, this mechanism takes advantage of the fast interrupt handling, which minimizes the overhead of MBR. 
Empirical results show that \textit{H-MBR} introduces no overhead for critical workloads, maintaining their performance and ensuring predictable behavior. For non-critical workloads, the mechanism introduces an overhead below 1\% for regulation periods equal or higher than 2 µs. It also fully eliminates critical workload performance degradation if strict budgets are enforced on non-critical workloads, demonstrating its reliability in MCSs.

\section{Interference and Interference Mitigation}
\label{sec:memory_utilization_background}

\begin{figure}[t]
    \centering
    \includegraphics[width=1\linewidth]{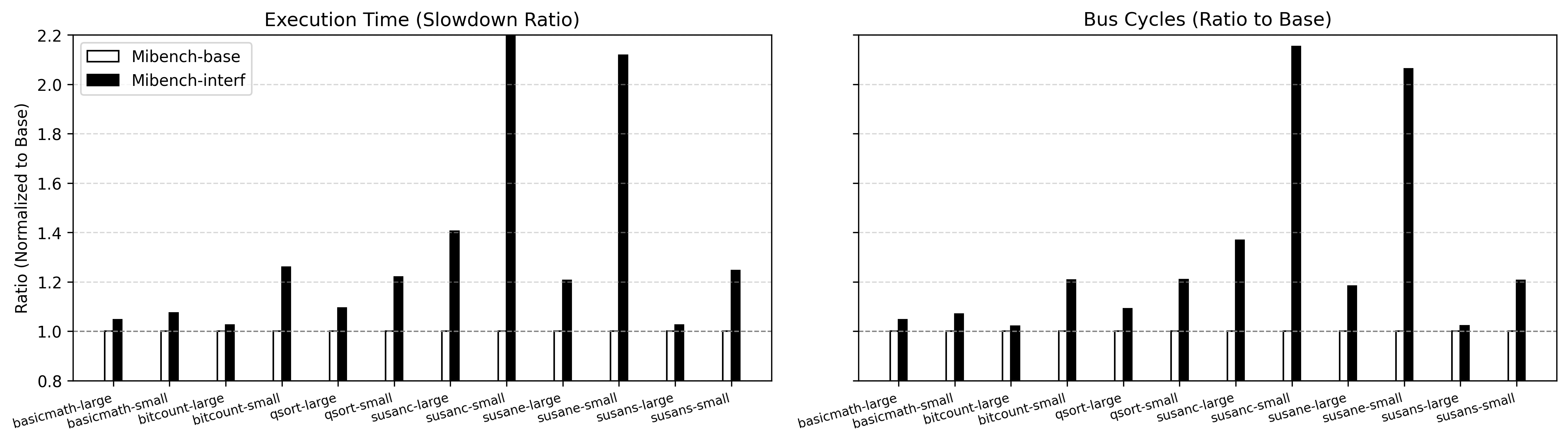}
    \caption{Mibench Execution Time and Bus Cycles (Ratio)}
    \label{fig:mibrnchbackground}
\end{figure}

In multicore embedded systems, contention generated in shared resources is a significant challenge that impacts performance, predictability, and temporal isolation, particularly in RT and MCS \cite{cazorla2019probabilistic, yun2014palloc, casini2020holistic}. 
Contention arises when multiple cores simultaneously access shared components~\cite{martins2020bao, serrano2021qos}, (i) such as main memory~\cite{yun2012memory, bandwatch, yun2013memguard, jailhouse, yun2014palloc}, (ii) last-level caches~\cite{martins2020bao, mancuso2013cache}, (iii) system buses~\cite{bandwatch}, and (iv) additional subsystems like the Generic Interrupt Controller~\cite{irqcoloring, irqcoloring2}. 
These shared resources can lead to delays as cores compete for access, resulting in increased response times and unpredictable behavior. This unpredictability is especially problematic in systems with strict timing requirements, where such delays affect the system’s temporal isolation, threatening reliability and safety~\cite{farshchi2020bru, mancuso2013cache, crespo2018feedback}. As shown in Figure \ref{fig:mibrnchbackground}, this interference can arise up to 2.3x on memory-intensive benchmarks like in Mibench's susanc-small. As system complexity grows, understanding and controlling memory utilization becomes essential to mitigate interference effects and ensure critical tasks' predictability.
To mitigate memory-induced contention and ensure temporal isolation of tasks, several techniques have been proposed.

\label{sec:cache_coloring_mbr}

\par \textbf{Cache Partitioning.} Cache partitioning divides LLC into distinct regions, or “partitions,” which are assigned to specific workloads to control cache access and reduce contention. Two primary approaches to cache partitioning are (i) cache locking and (ii) cache coloring. Cache locking relies on hardware assistance to restrict eviction from designated cache lines, securing specific portions of the cache for high-priority tasks. Cache coloring, on the other hand, uses the overlap between virtual page numbers and cache indices to partition cache sets without needing specialized hardware. Hybrid techniques, such as Colored Lockdown \cite{mancuso2013cache}, combine coloring and locking, while other approaches propose dynamic re-coloring schemes to adapt to varying workloads \cite{ye2014coloris, xu2017vcat, roozkhosh2020bleaching}. Cache coloring has been successfully implemented in hypervisors like Bao \cite{martins2020bao}, Jailhouse \cite{jailhouse}, XVisor \cite{modica2018hypervisor}, Xen \cite{xencoloring} and KVM \cite{kvmcoloring}, effectively enhancing predictability in multicore systems with real-time requirements. However, the efficiency of cache coloring is workload-dependent: for high-demand cores, it significantly reduces cache interference and improves performance, but with lighter or variable loads, it may underutilize cache space, potentially impacting overall system efficiency. Additionally, cache coloring is hardware-specific, requiring an understanding of cache structure details such as size, associativity, and mapping policies, which can limit its portability across platforms.

\definecolor{grayblue}{RGB}{180, 200, 220}
\definecolor{lightcolor}{RGB}{220, 235, 245}

\begin{table}[t]
\centering
\resizebox{\linewidth}{!}{ 
\begin{tabular}{>{\centering\arraybackslash}p{3.5cm} >{\centering\arraybackslash}p{1.5cm} >{\centering\arraybackslash}p{1.5cm} >{\centering\arraybackslash}p{3.5cm} >{\centering\arraybackslash}p{1.5cm} >{\centering\arraybackslash}p{2cm}}
\toprule
\rowcolor{grayblue} \textbf{Paper} & \textbf{Dynamic} & \textbf{Static} & \textbf{Implementation Level} & \textbf{CPU-centric} & \textbf{VM-centric} \\ 
\midrule
\rowcolor{lightcolor} A. Zuepke el al. \cite{mempol} & \CIRCLE & \Circle & Firmware & \CIRCLE & \Circle \\ 
\rowcolor{white} H.Yun et al.\cite{yun2013memguard} & \CIRCLE & \Circle & OS(Linux Kernel) & \CIRCLE & \Circle \\ 
\rowcolor{lightcolor} H.Yun el al. \cite{memguard2} & \CIRCLE & \Circle & OS(Linux Kernel) & \CIRCLE & \Circle \\ 
\rowcolor{white}  H.Yun et al.\cite{yun2014palloc} & \CIRCLE & \Circle & OS(Linux Kernel) & \CIRCLE & \Circle \\ 
\rowcolor{lightcolor} P. Sohal et al.\cite{Sohal2022} & \CIRCLE & \Circle & Hardware & \CIRCLE & \Circle \\ 
\rowcolor{white} A. Agrawal et al.\cite{othermancuso} & \CIRCLE & \Circle & OS(Linux Kernel) & \CIRCLE & \Circle \\ 
\rowcolor{lightcolor} E. Seals \cite{bandwatch} & \CIRCLE & \Circle & OS(Linux Kernel) & \CIRCLE & \Circle \\ 
\rowcolor{white} D. Hoornaert et al. \cite{memory} & \CIRCLE & \Circle & Hypervisor and Hardware & \CIRCLE & \Circle \\ 
\rowcolor{lightcolor} M. Pagani et al. \cite{biondioutravez} & \CIRCLE & \Circle & Hardware & \CIRCLE & \Circle \\ 
\rowcolor{white} M. Xu et al.\cite{hypervisor2} & \CIRCLE & \Circle & Hypervisor & \CIRCLE & \Circle \\ 
\rowcolor{lightcolor} P. Modica et al.\cite{modica2018hypervisor} & \Circle & \CIRCLE & Hypervisor & \CIRCLE & \Circle \\ 
\rowcolor{white} E. Gomes et al.\cite{everaldo} & \CIRCLE & \Circle & Hypervisor & \CIRCLE & \Circle \\ 
\rowcolor{lightcolor} G. Brilli et al.\cite{jailhouseMBR} & \CIRCLE & \Circle & Hypervisor & \CIRCLE & \Circle \\ 
\midrule
\rowcolor{white} H-MBR & \Circle & \CIRCLE & Hypervisor & \Circle & \CIRCLE \\ 
\bottomrule
\end{tabular}
}
\caption{Gap analysis}
\label{tab:memory}
\end{table}

\par \textbf{Memory Bandwidth Reservation (MBR).} MBR is a critical approach to reserving memory bandwidth, designed to prevent hardware contention on memory accesses, which is vital for achieving predictable performance in MCS \cite{farshchi2020bru, yun2014palloc}. When memory is heavily utilized, increased memory contention leads to bus delays, subsequently extending execution times. MBR works by reserving memory bandwidth quotas per core or VM, either statically or dynamically, effectively reducing contention on shared resources and improving overall performance \cite{cazorla2019probabilistic, modica2018hypervisor}. Most MBR implementations operate at the OS level \cite{yun2013memguard, mancuso2013cache, yun2014palloc, casini2020holistic, serrano2021qos, farshchi2020bru, zini2022io, bechtel2019cache} particularly within the Linux Kernel, where they manage bandwidth based on task requirements to ensure that critical tasks receive adequate access without overloading the memory controller \cite{casini2020holistic, serrano2021qos}. However, this OS-level integration limits MBR’s portability to other OSs, such as RTOSs like Zephyr and FreeRTOS, which are especially prevalent in real-time applications \cite{guthaus2001mibench, pinto2019trustzone}. On the other hand, some MBR implementations use FPGA-dedicated accelerators \cite{zini2022io} to monitor and control memory bandwidth, which has the significant downside of being platform-specific. Additionally, some approaches implement MBR at the hypervisor level \cite{modica2018hypervisor, hypervisor2, everaldo, jailhouseMBR}; however, existing methods are applied at the vCPU level and are integrated into hypervisors with slower interrupt handling compared to Bao~\cite{martins2023static}. This slower response introduces significant overhead, particularly for time-sensitive applications, where Bao's efficient, low-latency interrupt handling offers a clear advantage. Additionally, some of this approaches are tied to specific architecture, reducing it's portability. Table~\ref{tab:memory} compares various MBR techniques, highlighting that most Linux OS-based implementations are dynamically managed yet tied to the Linux kernel, restricting their flexibility and scalability across diverse embedded environments.

\par \textbf{Gap Analysis.}  MemGuard \cite{yun2013memguard} presents a dynamic allocation in Linux kernel while H.Yun et al. \cite{memguard2} presents an extension to previous implementation. Palloc\cite{yun2014palloc} implements a bank-aware memory allocation strategy. A. Zuepke et al. \cite{mempol} adopt a distinct methodology by utilizing an MCU to control a larger APU. A. Agrawal et al. \cite{othermancuso}, and E. Seals \cite{bandwatch} all implement dynamic allocation within the Linux kernel. Hardware-level implementations come from P. Sohal et al.\cite{Sohal2022} and M. Pagani et al.\cite{biondioutravez}, both using dynamic allocation. At the hypervisor level, E. Gomes et al. \cite{everaldo} and G. Brilli et al. \cite{jailhouseMBR} leverage MemGuard on the hypervisor level, focusing on CPU-centric approaches. Moreover, M. Xu et al.\cite{hypervisor2} implement dynamic allocation, while P. Modica et al.\cite{modica2018hypervisor} opt for static allocation. D. Hoornaert et al. \cite{memory} uniquely combine hypervisor and hardware approaches with dynamic allocation. While all these solutions focus on CPU-centric approaches, \textit{H-MBR} introduces a VM-centric approach that aligns with MCSs' perspective by enabling bandwidth allocation based on VMs than individual cores, implementing static allocation at the hypervisor level.

\section{The mechanism}
\begin{figure}
    \centering
    \includegraphics[width=0.98\linewidth]{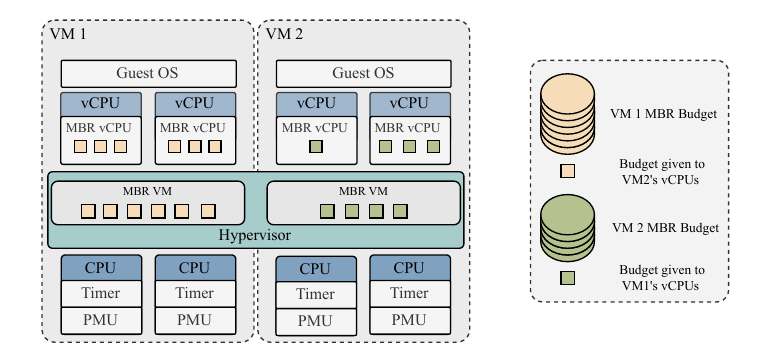}
    \caption{System Overview}
    \label{fig:systemoverview}
\end{figure}

\subsection*{H-MBR Interface}

\par \textit{H-MBR} focuses on a VM-centric perspective, where each VM requires two key MBR parameters: (i) \textit{budget} and (ii) \textit{period} configured at compile time. The MBR configuration operates on a per-VM basis, enabling customized budgets and regulation periods for individual VMs. Additionally, the VM budget distribution across CPUs can be handled in two distinct ways: through automatic balancing or via non-balanced distribution, where specific allocations can be manually set. This flexibility allows for uneven distribution of the VM budget across its vCPUs, as demonstrated by the green-colored VM in Figure \ref{fig:systemoverview}.

\subsection*{H-MBR Run-Time}

\begin{algorithm}[b]
\caption{Initial Assignment of MBR Parameters}
\label{alg:vm_mbr_assignment}
\begin{algorithmic}[1]
\FOR{each VM in VMs}
    \STATE \texttt{VM.budget} $\leftarrow$ \texttt{budget\_vm}
    \STATE \texttt{VM.period} $\leftarrow$ \texttt{period\_vm}
    \FOR{each vCPU in VM.vCPUs}
        \STATE \texttt{vCPU.budget} $\leftarrow$ \texttt{VM.has\_custom\_dist}
        \STATE \hspace{1cm} \texttt{? (VM.budget} $\times$ \texttt{vCPU\_percentage[vCPU])}
        \STATE \hspace{1cm} \texttt{: (VM.budget / VM.num\_vCPUs)}
    \ENDFOR
\ENDFOR
\end{algorithmic}
\end{algorithm}

\begin{figure}[t]
    \centering
    \begin{minipage}[t]{0.49\linewidth}
        \centering
        \includegraphics[width=\linewidth]{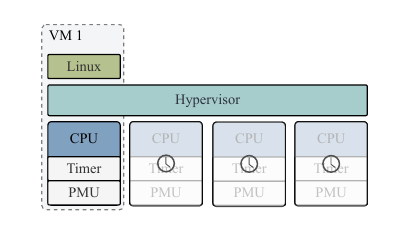}
        \caption{ Standalone setup with single Linux VM.}
        \label{fig:solo_setup}
    \end{minipage}
    \hfill
    \begin{minipage}[t]{0.49\linewidth}
        \centering
        \includegraphics[width=\linewidth]{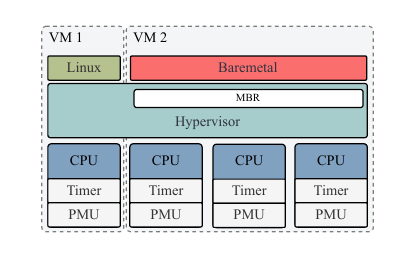}
        \caption{Co-existing setup with two VMs: (i) a Linux VM and (ii) a baremetal VM.}
        \label{fig:interf_setup}
    \end{minipage}
\end{figure}

The MBR mechanism follows a Memguard \cite{yun2013memguard} based approach on how to track this metrics in real-time, thereby the mechanism relies on two key peripherals: (i) PMU and (ii) generic timer, respectively. Both of this peripherals exist physically inside each CPU, as depicted in figures \ref{fig:solo_setup} and \ref{fig:interf_setup}. The hypervisor-level approach leverages this physical peripherals common to most architectures and assigns them directly to each vCPU, which leads to CPU monitoring not impacting  other CPUs performance and therefore ensuring spatial and temporal isolation.   

\par \textbf{Timer.} The timer is configured to overflow after a defined \textit{Period}, setting the interval at which the memory bandwidth reserved for each core is reset. This periodic reset ensures that all cores start each \textit{period} with a reset \textit{budget}. Additionally, the timer re-activates any cores that were previously idled due to over-budget usage, allowing them to resume operation at the beginning of each new period. The overhead introduced by the timer interrupt can be quantified as follows:
{\(\text{Overhead}_{\text{timer}} = \frac{D_{\text{timer}}}{\text{Period}_{\text{timer}}}\)}, where \(D_{\text{timer}}\) corresponds to the total execution time of the timer interrupt (i.e., the sum of interrupt injection latency with the interrupt callback execution time).

\par \textbf{PMU.} The PMU tracks memory access activities by counting \textit{bus access} and triggers an overflow interrupt if a core exceeds its reserved \textit{Budget} within the \textit{Period}. When this overflow occurs, the PMU signals the MBR mechanism to temporarily idle the over-budget core. This action prevents excessive memory contention, helping to maintain consistent response times for critical tasks on other cores.

Furthermore, the MBR configuration which operates on a per-VM basis, enables customized budgets and regulation periods for individual VMs. The VM budget distribution across CPUs can be handled in two distinct ways: through automatic balancing or via non-balanced distribution, where specific allocations can be manually set.


\section{Evaluation}
This section provides a detailed explanation of the methodology used to obtain the results, including descriptions of the guest OSs, configurations, and setups on the target hardware. Each component was carefully chosen to analyze memory interference under varying conditions. 

\par \textbf{Target Platform \& Measurement Tools:} 
We conducted our experiments assessment on a  Zynq UltraScale+ ZCU104 board (ARMv8-A-based architecture), which features a quad-core Arm Cortex-A53 processor. Each CPU has a private cache (data and instruction, 32KiB each), and the unified shared L2 cache (1MiB). The Cortex-A53 processor also features a ARM PMUv3, which has been leveraged to profile benchmark execution and gather key microarchitectural events. The selected events include bus cycles and execution cycles, providing insight into memory and system bus usage. Additionally, we used the \texttt{perf} tool on the Linux OS as an interface to the CPU's PMU.

\par \textbf{VM Workloads:} We deployed two distinct guest environments for benchmarking: (i) a Linux-based VM, which will be denominated as Critical VM (C) and (ii) a baremetal VM, which will be denoted as Non-Critical VM (NC). 
For the critical VM, we deployed a Linux-based guest to enable the deployment of MiBench automotive suite \cite{guthaus2001mibench}, a widely-used benchmark for automotive embedded systems—to simulate real-world automotive workloads.
For the NC VM, we deployed a baremetal guest that continuously writes to a buffer sized to match the LLC capacity, creating intentional memory contention to thoroughly assess interference effects on the memory hierarchy. We focused exclusively on write operations, rather than reads or a combination of both, since write operations stresses even more the memory shared hardware resources, when compared to read operations. 


\par \textbf{Setups and Configurations:} 
We evaluated four distinct setups: (i) \textit{solo}, (ii) \textit{interf}, and (iii)\textit{interf+mbr}. The \textit{solo} configuration involves running Linux alone, whereas \textit{interf} includes both Linux and Baremetal guests to introduce memory interference. 
The suffix \textit{mbr} is used to identify setups with memory bandwidth reservation.
As for hardware resources, we followed the CPU assignment identified in Figure~\ref{fig:interf_setup}: one CPU is allocated to Linux in all configurations, and three CPUs are assigned to the baremetal VM, when included. 



\begin{figure}[t]
    \centering
    \begin{minipage}[t]{0.48\linewidth}
        \centering
        \includegraphics[width=\linewidth]{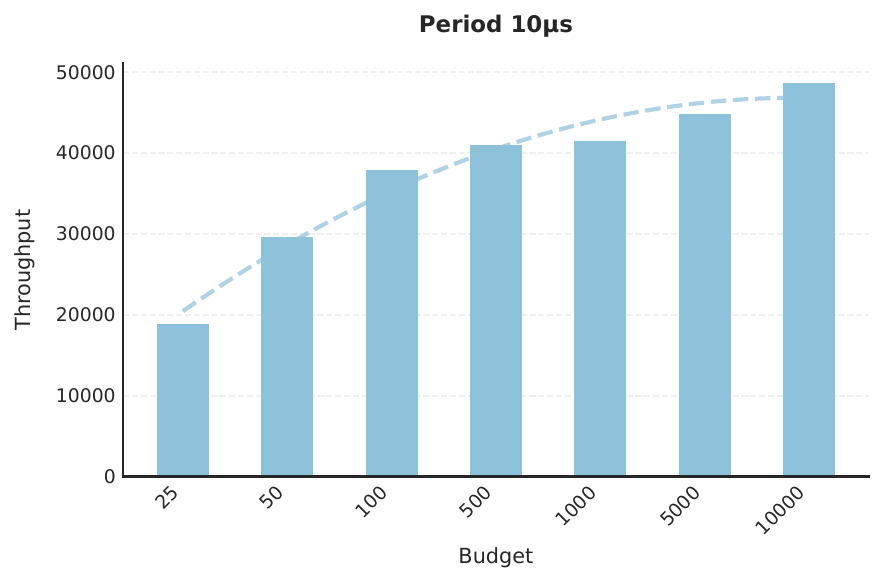}
        \caption{\textit{interf+mbr} NC throughput budget variation}
        \label{fig:throughput_config_a}
    \end{minipage}
    \hfill
    \begin{minipage}[t]{0.48\linewidth}
        \centering
        \includegraphics[width=\linewidth]{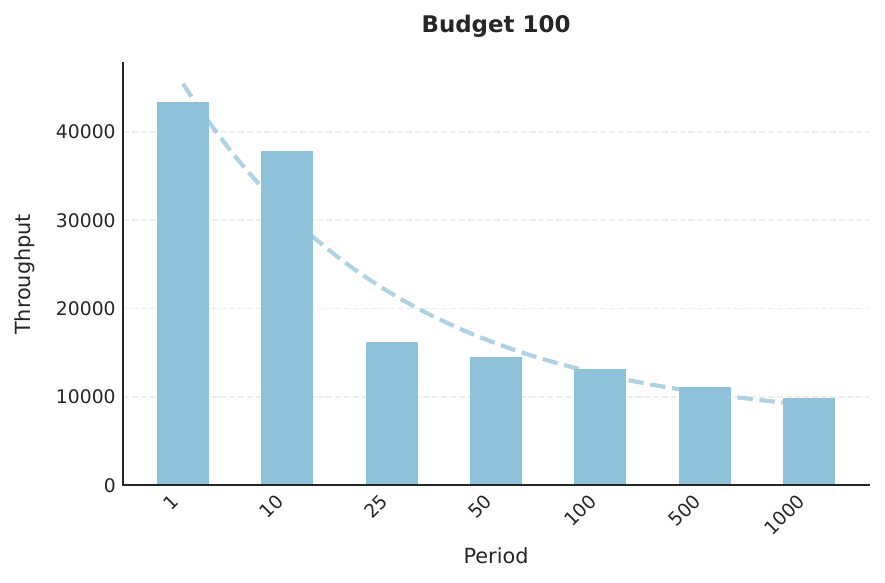}
        \caption{\textit{interf+mbr} NC throughput period variation.}
        \label{fig:throughput_config_b}
    \end{minipage}
\end{figure}
\subsection{Impact of MBR on Baremetal Throughput and Overhead} \label{sec:impactmbr}

To assess the effect of MBR configurations on the non-critical baremetal VM, we measured the number of cache line writes achievable under different budget and period settings. This section will focus on the memory-intensive \textit{susan-c} benchmark from the MiBench suite(the results of other benchmarks can be found in Appendix A) on the critical Linux-based VM. The \textit{susan-c} benchmark was specifically chosen for its high memory contention characteristics, making it an ideal workload for evaluating the impact of MBR adjustments on memory access behavior in the non-critical VM. 

\textbf{Budget Configuration.} Increasing the MBR period while keeping the budget fixed at 100 reduces the cache line write capacity of the non-critical, as shown in the Figure \ref{fig:throughput_config_a}. For instance, with a period of 25 µs, cache line writes decrease by 2.6 times compared to a period of 1 µs. This outcome reflects how longer periods effectively limit the NC’s access to memory bandwidth, as the delay in budget reset constrains the number of operations the non-critical can perform within each interval. These findings align with our expectation that increasing the period value reduces memory bandwidth availability for the non-critical, thus lowering its potential to interfere with the critical VM.

\textbf{Period Configuration.} Figure \ref{fig:throughput_config_b} demonstrates that with a fixed period of 10 µs, increasing the budget significantly increases the non-critical's cache line writes. For instance, a budget setting of 500 allows 2.2 times more cache line writes compared to a budget of 25. Higher budgets allocate more memory bandwidth to the non-critical VM, which in turn elevates its cumulative memory access capacity over time, maximizing the interference potential with the critical VM.

\begin{figure}[t]
    \centering
    \begin{minipage}[t]{0.48\linewidth}
        \centering
        \includegraphics[width=\linewidth]{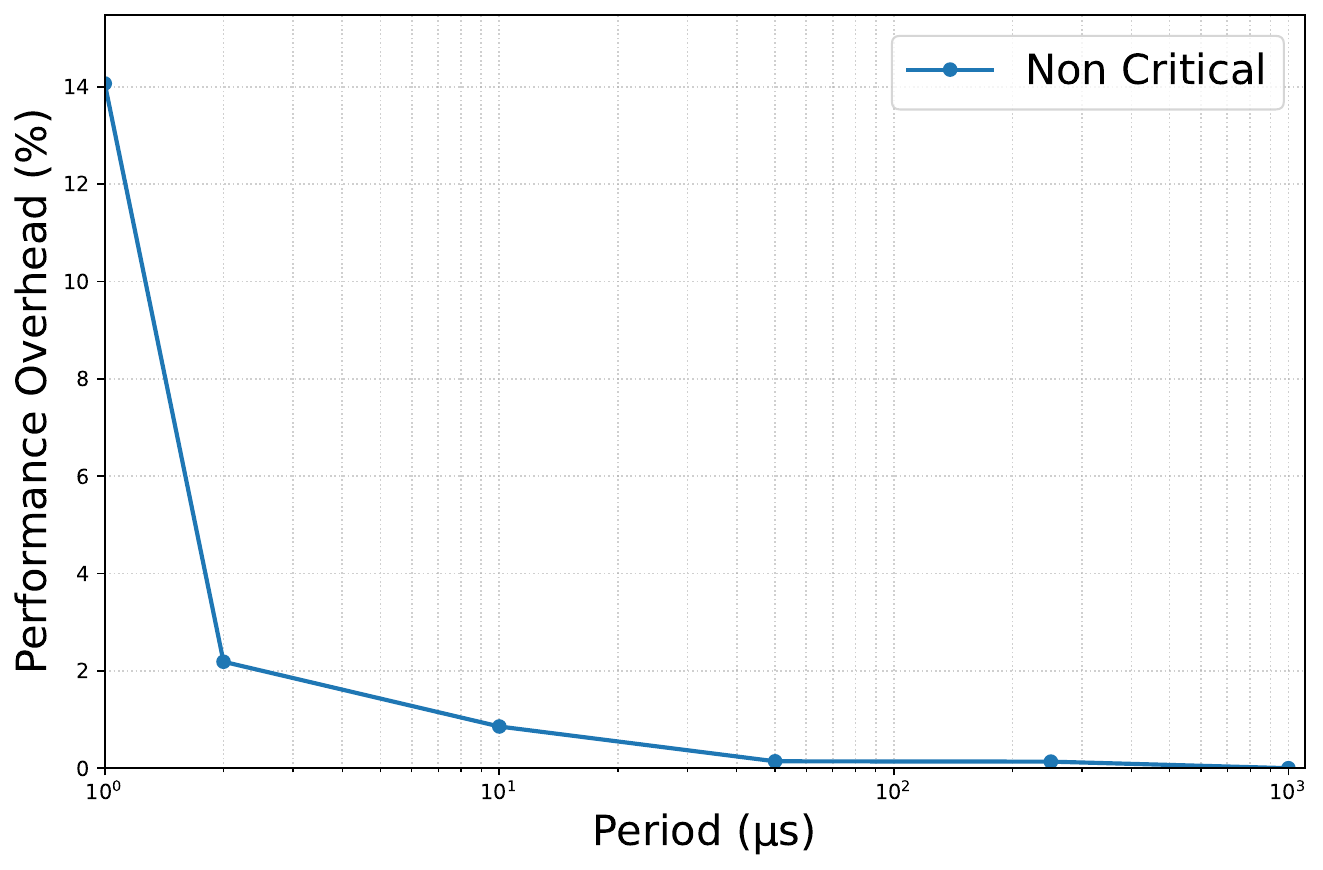}
        \caption{Performance overhead on NC.}
        \label{fig:performance_overhead}
    \end{minipage}
    \hfill
    \begin{minipage}[t]{0.48\linewidth}
        \centering
        \includegraphics[width=\linewidth]{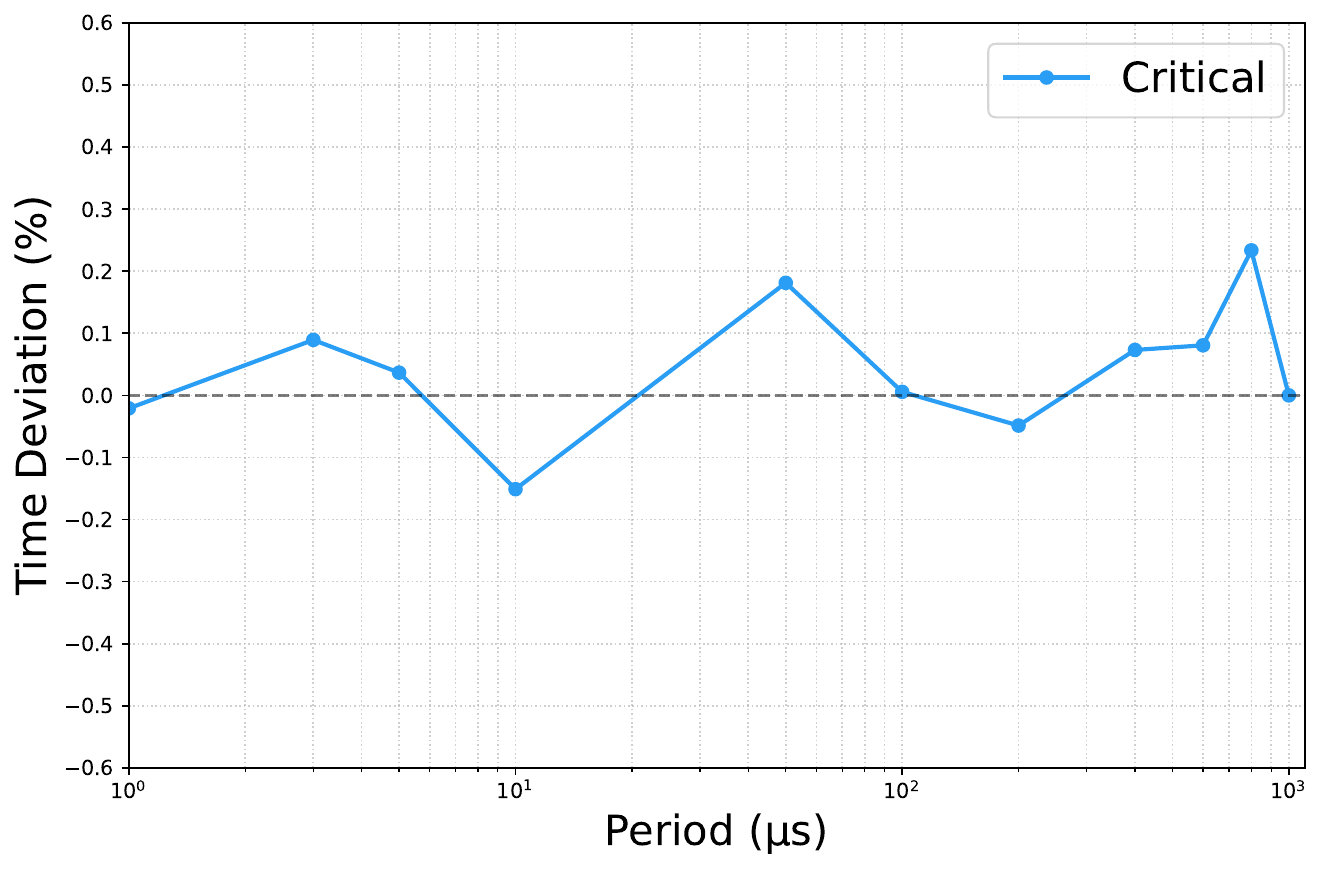}
        \caption{Time deviation on C.}
        \label{fig:time_deviation}
    \end{minipage}
\end{figure}

\textbf{Overhead.} Additionally, our empirical results show that \textit{H-MBR} provides VM-level isolation with minimal overhead. To accurately assess the MBR mechanism’s overhead, we deployed the baremetal VM in a solo setup, without the critical Linux VM. In this configuration, the PMU interrupt was disabled to prevent the baremetal from idling, as the PMU interrupt would otherwise trigger a negligible overhead due to the long idle wait times (which greatly exceed the combined interrupt latency and callback execution time). Consequently, the observed MBR overhead reflects a scenario where the budget never expires, and only the timer interrupt is active.
As shown in Figure \ref{fig:performance_overhead}, the observed overhead for the non-critical VM remains minimal overall, with a noticeable spike only at a regulation period of 1 µs, where it reaches 14.3\%. This elevated overhead is anticipated due to the frequent timer interrupts within such a short interval. However, as the regulation period increases slightly, the overhead rapidly diminishes: at 2 µs, it drops to 2\%, and by 10 µs, it is already below 1\%. For periods beyond 10 µs, the overhead becomes negligible, approaching zero. This trend demonstrates that the MBR mechanism maintains low interference impact on the non-critical VM, especially at moderate to higher regulation periods, reinforcing its effectiveness in providing controlled memory isolation.

\subsection{Effect of MBR on Linux Benchmark Performance}

This section presents results for the \textit{interf+mbr} scenario, where MBR is applied to the non-critical baremetal VM, leaving the critical Linux VM unregulated. This approach enables a focused evaluation of MBR’s impact on the critical VM's performance stability under varying MBR configurations. Two key aspects of MBR are analyzed: the influence of different budget values at a fixed period and the effect of varying periods at a fixed budget. These parameter variations directly affect memory bandwidth according to  $\text{Bandwidth} = \frac{\text{Budget}}{\text{Period}}$.

\begin{figure}
        \centering
    \includegraphics[width=1\linewidth]{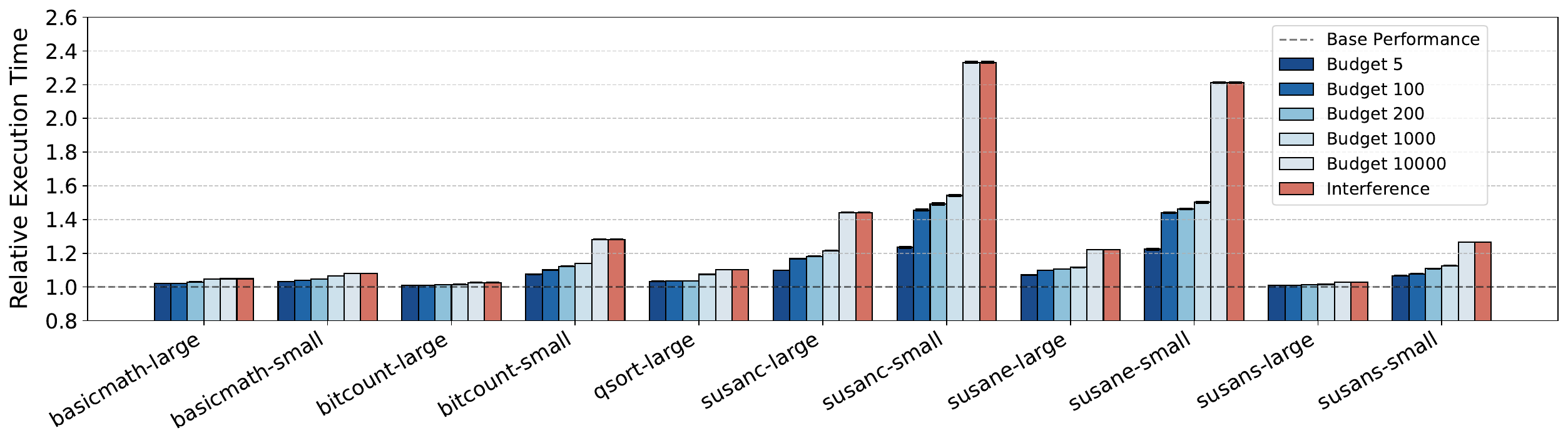}
    \caption{Fixed 10µs period. }
    \label{fig:Cfixedperiod}
\end{figure}

\textbf{Fixed Period, Variable Budgets.}
Figure \ref{fig:Cfixedperiod} presents the effects of increasing the MBR budget for the non-critical VM with a fixed 10 µs period. Results show that as the MBR budget allocated to the non-critical VM increases, performance degradation in the critical VM also rises across all benchmarked workloads. This outcome aligns with expectations: a larger MBR budget allows the non-critical baremetal application increased memory bandwidth, thereby intensifying interference within shared memory resources.
With budgets ranging from 50 to 10,000 memory accesses per period, lower budgets (e.g., a budget of 50) yield minimal interference, achieving performance levels close to the baseline for the critical VM. Higher budgets, such as 1,000 and 10,000, produce more substantial interference, notably in memory-intensive benchmarks like \textit{susanc} and \textit{susane}, where the relative execution time arises up to 2.3x and 2.2x, respectively. These benchmarks experience considerable performance degradation at higher budgets due to their sensitivity to memory contention. Interestingly, even with the short 10 µs period, small budgets (e.g., 100) still cause a measurable impact on the critical VM’s performance, increasing the execution time by 1.45x for the susanc-small\_benchmark. 

\begin{figure}
        \centering
        \includegraphics[width=1\linewidth]{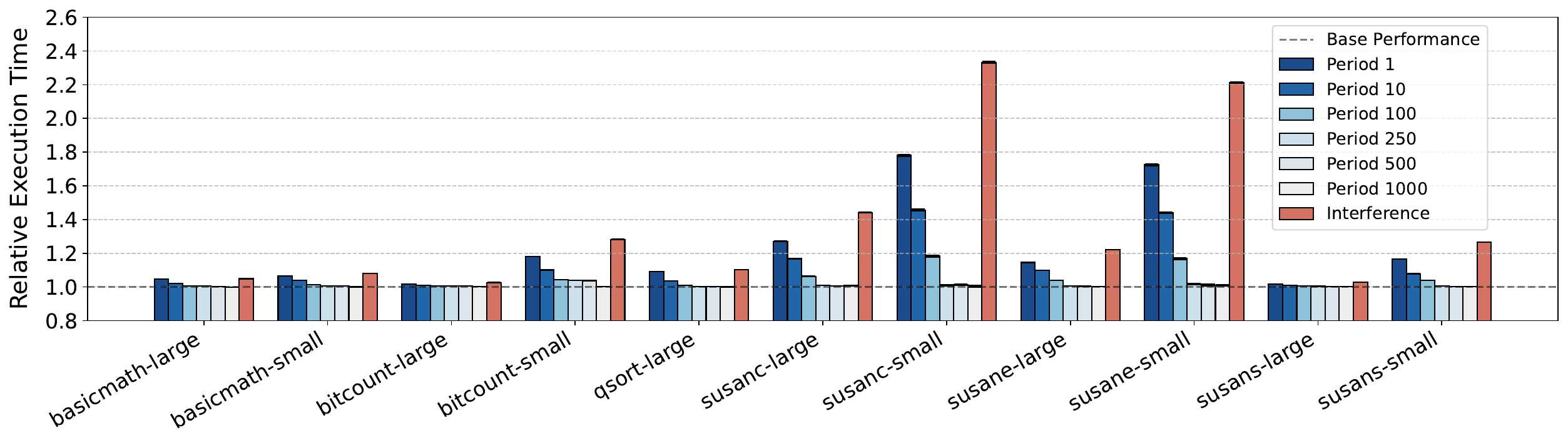}
        \caption{Fixed budget of 100}
        \label{fig:Cfixedbudget}
\end{figure}

\textbf{Fixed Budget, Variable Periods.} 
Figure \ref{fig:Cfixedbudget} illustrates how different regulation periods, from 1 µs to 1000 µs, impact critical VM performance when the MBR budget for the non-critical VM is fixed at 100. Here, we observe that longer periods provide greater protection to the critical VM by reducing memory contention. This is especially evident in memory-sensitive applications, where the extended periods give the critical VM uninterrupted access to memory resources before the non-critical VM’s budget resets, thus improving performance. For instance, with a 1 µs period, critical benchmarks like \textit{susanc} show a performance increase of 1.78x compared to the solo execution. However, as the period extends to 1000 µs, performance approaches solo levels, indicating a substantial reduction in interference. These results underscore that longer periods limit the non-critical VM’s memory bandwidth utilization, effectively mitigating its impact on the critical VM’s workloads.

\textbf{}

\textbf{Overhead.} The results presented in Figure \ref{fig:time_deviation} confirm that applying MBR on the non-critical VM does not introduce any overhead on the critical VM, as each VM has dedicated, isolated CPUs. Additionally, the timer interrupts from MBR that regulate the non-critical VM’s memory access do not interfere with the critical VM, resulting in effectively zero overhead. This is evidenced by the consistency in the critical VM’s execution times across various MBR regulation periods, where variations remain minimal, within the range of 0.01\% to 0.2\%, equating to a minor fluctuation of around 20 µs. These findings underscore MBR’s effectiveness for mixed-criticality systems, as it maintains reliable performance in the critical VM without impact from adjustments in the non-critical VM.

\section{Discussion and Future Work}
 Currently, the MBR budget and period parameters are set statically at compile-time. Future work encompasses the introduction of dynamic configuration based on runtime monitoring of memory usage patterns could enable more adaptive and optimized bandwidth allocation. This could involve leveraging machine learning techniques to predict memory demands and proactively adjust MBR settings.
Integration with additional isolation mechanisms, such as cache coloring, is another promising direction. By combining MBR with these techniques, contention could be mitigated at multiple levels of the memory hierarchy, providing even stronger temporal isolation guarantees. The interplay between these mechanisms and their cumulative impact on performance and predictability warrants further investigation.
Benchmarking other guest OSs with MBR is also an important consideration. Since the mechanism targets real-time applications, and already supports other OSs, benchmarking a RTOS like Zephyr, would further enhance \textit{H-MBR's} potential. Demonstrating the portability and generalizability of \textit{H-MBR} would broaden its applicability across a wider range of use cases.
Finally, long-term efforts could explore extending \textit{H-MBR} beyond CPUs to heterogeneous computing elements like GPUs, FPGAs, and AI accelerators. As embedded systems increasingly incorporate these specialized components, managing their shared memory resources becomes critical. Adapting MBR mechanisms to these contexts could unlock new possibilities for predictable acceleration in mixed-criticality environments.

\section{Conclusion}

This work introduced \textit{H-MBR}, a VM-centric MBR mechanism for MCS on multicore platforms. Through extensive evaluations, the mechanism demonstrated its effectiveness in mitigating memory contention and enhancing isolation between VMs, while it presented a remarkably low overhead. The evaluation results showed that \textit{H-MBR }significantly reduced the interference on critical tasks by carefully controlling memory bandwidth reservation. Under configurations where MBR budget was kept low, critical workloads showed minimal to no performance degradation. Overhead analysis further confirmed that \textit{H-MBR} imposes minimal impact on workloads, even when dealing with minor periods. In the worst-case scenario of a 1 µs regulation period, the mechanism incurs an overhead of up to 14\%; however, for regulation periods of 2 µs or longer, this overhead drops to below 1\%. Compared to existing solutions, this is the lowest interrupt overhead observed due to Bao’s optimized interrupt handling and lightweight design. This capability makes \textit{H-MBR} stand apart from other implementations, specially in embedded applications, where it maximizes available processing resources for critical tasks without compromising real-time performance.

\bibliography{oasics-v2021-sample-article}

\appendix \label{appendix:benchmarks}

\section{Appendix}

\section*{Impact of MBR on Baremetal Throughput and Overhead}




This appendix builds on the discussion in Section 4.1, which evaluated the memory-intensive \textit{susan-c} benchmark running on the critical Linux-based VM (C) and focused on the throughput of the non-critical VM (NC). Here, we extend the analysis to additional benchmarks—\textit{susan-e},\textit{ susan-s}, \textit{basicmath}, \textit{qsort}, and \textit{bitcount}—under the same conditions to provide a broader perspective on MBR's impact on different workloads. By examining these diverse benchmarks, we gain deeper insights into the effectiveness of MBR in managing memory contention across different types of computational tasks.

\begin{figure}[h]
   \centering
   \begin{subfigure}[b]{0.3\textwidth}
       \includegraphics[width=\textwidth]{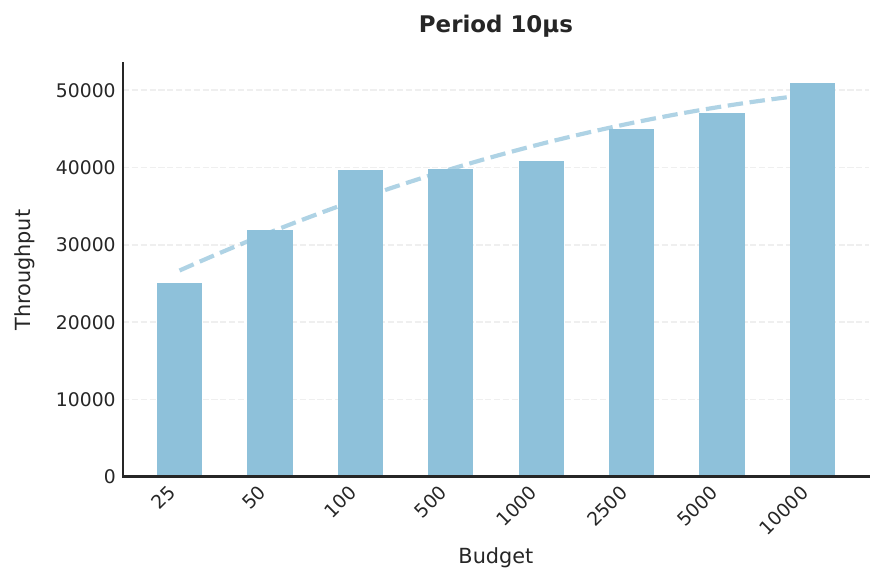}
       \caption{\textit{susan-e small}}
       \label{fig:susan_e}
   \end{subfigure}
   \hfill
   \begin{subfigure}[b]{0.3\textwidth}
       \includegraphics[width=\textwidth]{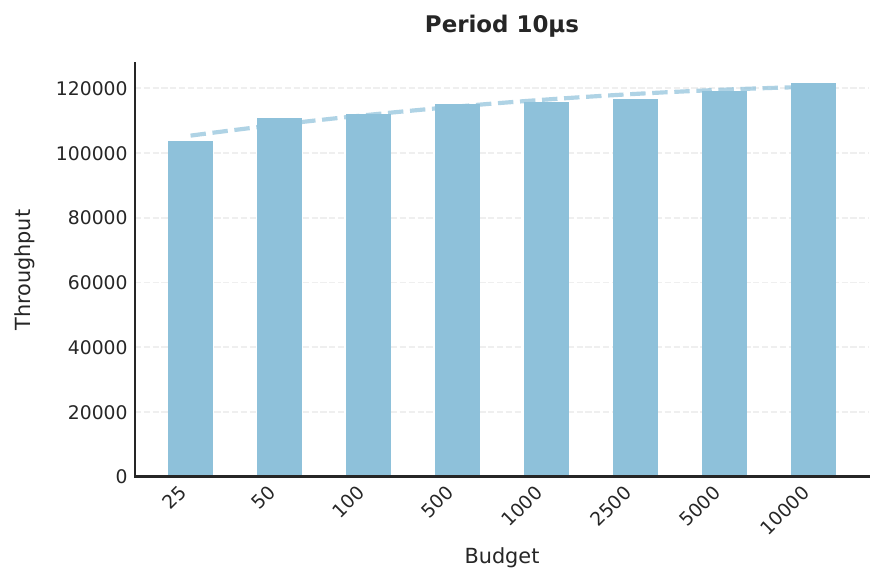}
       \caption{\textit{susan-s small benchmark}}
       \label{fig:susan_s}
   \end{subfigure}
   \hfill
   \begin{subfigure}[b]{0.3\textwidth}
       \includegraphics[width=\textwidth]{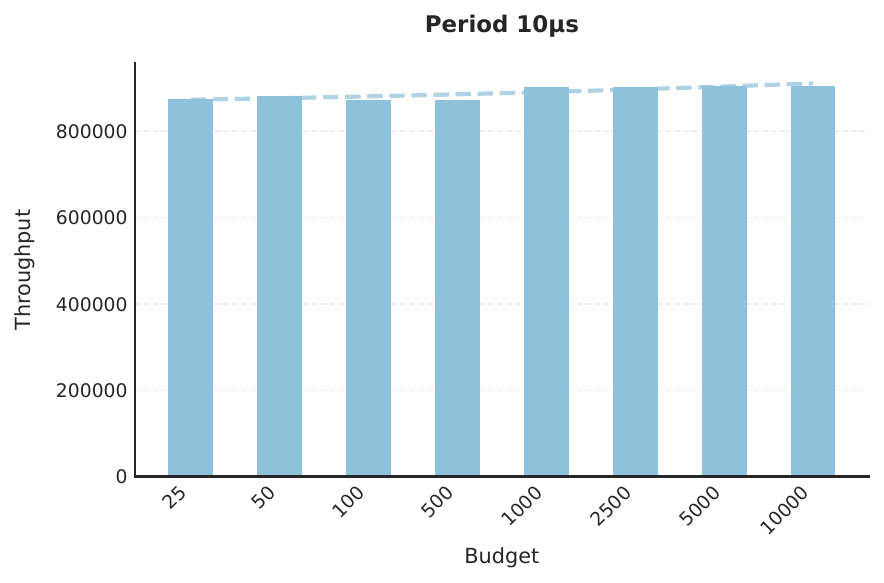}
       \caption{\textit{basicmath} small benchmark}
       \label{fig:basicmath}
   \end{subfigure}

   \vspace{0.5cm} 

   \begin{subfigure}[b]{0.3\textwidth}
       \includegraphics[width=\textwidth]{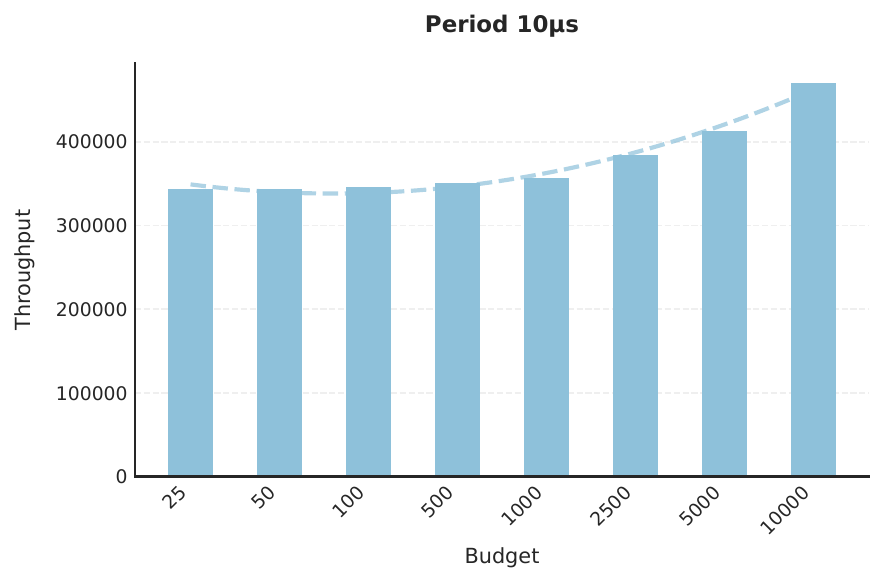}
       \caption{\textit{qsort} small benchmark}
       \label{fig:qsort}
   \end{subfigure}
   \hfill
   \begin{subfigure}[b]{0.3\textwidth}
       \includegraphics[width=\textwidth]{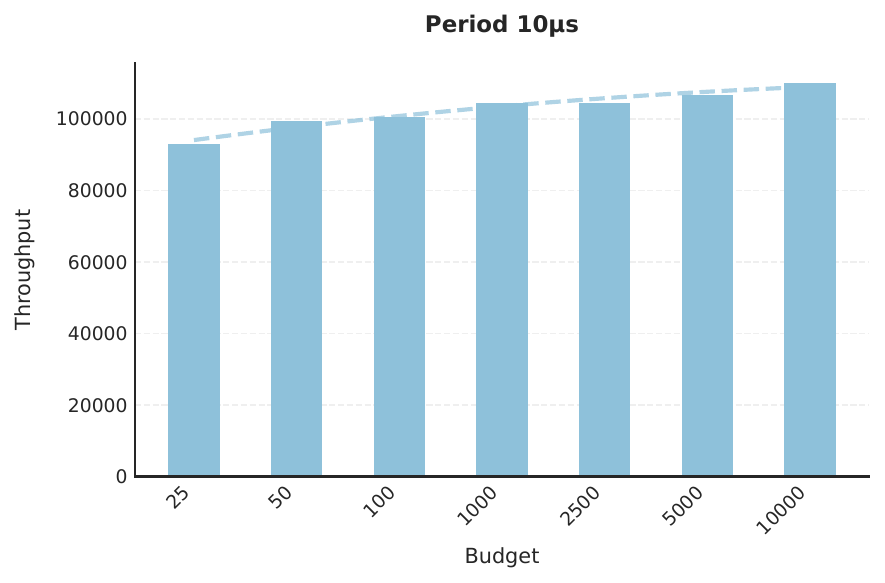}
       \caption{\textit{bitcount} small benchmark}
       \label{fig:bitcount}
   \end{subfigure}
   \hspace{0.3\textwidth} 

   \caption{Throughput variation on NC VM with different budget configurations for various benchmarks. Benchmarks exhibit different levels of sensitivity to budget changes based on their memory intensity.}
   \label{fig:all_benchmarks}
\end{figure}

    \begin{itemize}
        \item \textbf{\textit{Susan-e and Susan-s (Figures \ref{fig:susan_e} and \ref{fig:susan_s}):}} The NC demonstrated significant improvements in throughput as the budget increased. These results underscore the high memory bandwidth usage of both benchmarks.
        \item \textbf{\textit{Basicmath (Figure \ref{fig:basicmath}):}} Unlike the other benchmarks, the NC running alongside \textit{basicmath} displayed minimal sensitivity to budget changes, maintaining consistently high throughput across all configurations. This stability underscores the benchmark's lower reliance on memory bandwidth for its computational tasks.
        \item \textbf{\textit{Qsort (Figure \ref{fig:qsort}):}} The NC VM showed moderate throughput improvement as the budget increased, though not as dramatically as with the \textit{susan} benchmarks, indicating moderate memory bandwidth dependency.
        \item \textbf{\textit{Bitcount (Figure \ref{fig:bitcount}):}} The NC VM's throughput showed moderate improvement with increasing budgets, reflecting the benchmark's medium memory bandwidth dependency, with a more gradual scaling compared to \textit{susan-e} and \textit{susan-s}.
        \end{itemize}

These findings underscore the adaptability and practical impact of MBR in managing memory bandwidth effectively. Memory-intensive benchmarks such as \textit{susan-e}, \textit{susan-s}, and \textit{bitcount} display noticeable improvements in NC throughput when allocated higher budgets, reflecting their heavy reliance on memory resources. Conversely, memory-lighter benchmarks like \textit{basicmath} remain largely unaffected, proving that they are less constrained by memory bandwidth. This demonstrates the importance of workload-specific MBR tuning: adjusting budgets based on workload demands can maximize system performance while preserving isolation. 

\end{document}